\documentclass[10pt,journal,a4paper]{IEEEtran}

\usepackage{cite}
\usepackage{subfig}
\usepackage[dvips]{graphicx}
\usepackage[latin1]{inputenc}
\usepackage{url}
\usepackage{amsfonts}
\usepackage{amsmath}
\usepackage{times}
\usepackage{algpseudocode}
\usepackage{algorithm}
\usepackage{enumerate}
\usepackage{siunitx}
\DeclareSIUnit \dBm {dBm}
\DeclareSIUnit \dB {dB}
\DeclareSIUnit \Mbps {Mbps}
\usepackage{multirow}
\usepackage{tikz}

\newtheorem{theorem}{Theorem}[section]
\newtheorem{corollary}{Corollary}[section]
\newtheorem{remark}{Remark}[section]

\newtheorem{proposition}{Proposition}[section]

\begin{document}

\title{Wireless Vehicular Networks in Emergencies:\\ A Single Frequency Network Approach}

\author{\IEEEauthorblockN{Andrea Tassi\IEEEauthorrefmark{1}, Malcolm Egan\IEEEauthorrefmark{3}, Robert J. Piechocki\IEEEauthorrefmark{1} and Andrew Nix\IEEEauthorrefmark{1}}

\IEEEauthorblockA{\IEEEauthorrefmark{1}Department of Electrical and Electronic Engineering, University of Bristol, United Kingdom}

\IEEEauthorblockA{\IEEEauthorrefmark{3}Laboratoire de Math{\'e}matiques, UMR 6620 CNRS, Universit{\'e} Blaise Pascal, Clermont-Ferrand, France}}

\maketitle

\begin{abstract}
Obtaining high quality sensor information is critical in vehicular emergencies. However, existing standards such as IEEE 802.11p/DSRC and LTE-A cannot support either the required data rates or the latency requirements. One solution to this problem is for municipalities to invest in dedicated base stations to ensure that drivers have the information they need to make safe decisions in or near accidents. In this paper we further propose that these municipality-owned base stations form a Single Frequency Network (SFN). In order to ensure that transmissions are reliable, we derive tight bounds on the outage probability when the SFN is overlaid on an existing cellular network. Using our bounds, we propose a transmission power allocation algorithm. We show that our power allocation model can reduce the total instantaneous SFN transmission power up to $20$ times compared to a static uniform power allocation solution, for the considered scenarios. The result is particularly important when base stations rely on an off-grid power source (i.e., batteries).
\end{abstract}


\section{Introduction}\label{sec:intro}
An important factor in vehicle-related accidents is a lack of information: if each driver is aware of their surroundings and road conditions, many accidents can be avoided~\cite{NHTSA}. As autonomous vehicles begin to gain traction, the problem of ensuring that each vehicle is properly informed will only be further exacerbated~\cite{C_ITS,5G-PPP}. A key question is therefore how high quality data can be shared by an Intelligent Transportation System (ITS) to aid drivers in emergency situations.

At present, there are no guarantees that existing \mbox{operator-owned} base station rollouts will be able to support communication between vehicles and transport authorities in emergencies. Moreover, vehicular communication standards such as IEEE 802.11p/DSRC can only realistically support data rates that barely exceed \SI{6}{\Mbps}~\cite{Kenney2011}. On the other hand, traditional 3GPP Long Term Evolution-Advanced (LTE-A) cellular deployments can achieve maximum data rates of \SI{100}{\Mbps} but have latencies of around \SI{100}{\milli\second}~\cite{Uhlemann2016}, which is not sufficient to ensure that each driver near or in an emergency can make informed decisions on routing or evasive maneuvers~\cite{GPMN16}.

These limitations of existing standards mean that in vehicular emergencies it is challenging to prevent increasing traffic density and an increase in the likelihood of additional accidents. For instance, it is difficult to support 5G-PPP's ``Bird's Eye'' use case, which corresponds to a road intersection monitored by a set of panoramic camcorders or Light Detection And Ranging (LiDAR) sensors. These sensors then broadcast live feeds at a bitrate of several megabit-per-second to approaching vehicles to notify drivers of possible hazards ahead \cite{C_ITS,5G-PPP}.

One solution is to install or switch on additional base stations dedicated to assisting in emergency situations. However, as the new base stations need to be regularly moved and re-installed for each emergency situation, or are utilized only for short time periods, there is little incentive for operators to own the infrastructure. As such, third parties including municipalities or local governments will need to intervene. To ensure that there is not a conflict between the operator-owned network and the emergency municipality-owned network, a network sharing agreement is required to ensure that the Quality of Service (QoS) requirements of all users are \mbox{met~\cite{7294664,7343930}}.

Currently, only a small number of network sharing arrangements to support dynamic instantiations of wireless networks in response to dynamic changes in demand (such as in emergencies) have been proposed. However, the networks without borders vision~\cite{Doyle2014} now provides a framework to support the development of novel sharing arrangements. In particular, third-party operated networks have been proposed for indoor network sharing in~\cite{Markendahl2013}, as well as between operators facilities such as power plants and large residential blocks in~\cite{Egan2015}.

In this paper, we propose that the municipality-owned network to aid in vehicle-related emergencies consists of a small number of base stations that operate as a Single Frequency Network (SFN)~\cite{sesia2011lte}. As such, multiple neighboring base stations (forming the SFN) broadcast the same Point-to-Multipoint (PtM) data streams in a synchronous fashion. This transmission mode has become increasingly common in \mbox{LTE-A} systems, where it is also known as the SFN-evolved Multimedia Broadcast and Multicast Service (eMBMS)~\cite{jsacTassi}. SFNs controlled by simplified core network architectures have already proved effective in vehicular communication systems, as they can ensure low end-to-end communication latencies and high bitraes~\cite{7452263,7115909,7390802}.

There are two key challenges that must be resolved in order for a realistic network sharing agreement with operators to be introduced~\cite{Doyle2014}. First, users served by the \mbox{operator-owned} network must not have a reduced QoS. To achieve this, the municipality-owned network can use highly directional antennas, with the main lobes aimed directly at emergency users. Second, emergency users must be served with a high Signal-to-Noise and Interference Ratio (SINR) to ensure that high quality data can be shared to inform the decisions of drivers near or in the emergency. To this end, we exploit the stochastic geometry framework to derive novel analytical guarantees on the SINR outage probability. These guarantees then form a basis for power control optimization to meet SINR targets while minimizing the power usage, which is particularly important when the base stations are installed only with batteries rather than a power source connected to the grid.

Stochastic geometry techniques now form an established methodology for characterizing wireless network performance \cite{haenggi2013stochastic}, where interfering base stations are modeled via a spatial Poisson point process (PPP). In this methodology, the SINR is treated as a random variable over the locations of the interfering base stations. There have only been limited attempts to treat SFNs using stochastic geometric techniques. In particular, Talarico~\textit{et al.}~\cite{6854956} considered multiple contiguous homogeneously distributed SFNs; however, this work did not consider any other base stations that might interfere with the SFN deployment, which are a key feature in our model.

Our key result is a new upper bound on the SINR outage probability, which is asymptotically tight as the radius of the operator-owned network increases. Moreover, we show that our upper bound is a good approximation for a city the size of London with realistic operator-owned base station densities. Numerical evaluation of our bound demonstrates that with a maximum SFN base station transmit power of only \SI{4.4}{\watt} a target outage probability of $0.1$ is possible with an SINR target of \SI{6.3}{\dB}, for a SFN comprising $3$ base stations located at a maximum distance of $\SI{300}{\meter}$ from an emergency user.

The remainder of the paper is organized as follows. In Section~\ref{sec:SM}, we detail our system model consisting of the municipality and operator-owned networks. In Section~\ref{sec:Res}, we derive our analytical guarantees on the outage probability and we use our analysis to optimize the power control. We numerically evaluate the network in Section~\ref{sec:NR} and conclude in Section~\ref{sec:CL}.

\section{System Model}\label{sec:SM}
\begin{figure}[t]
	\centering
	\includegraphics[width=0.9\columnwidth]{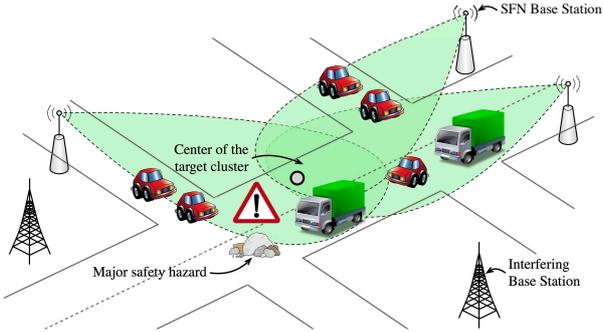}
	\caption{Considered system model.}
	\label{fig.f_1}
\end{figure}

Consider a municipality-owned SFN that provides emergency coverage to a small area of a city (illustrated in Fig.~\ref{fig.f_1}). In particular, the SFN serves a target cluster of vehicles to ensure that each vehicle can reliably receive information to support improve road safety; for example, through a Bird's Eye view~\cite{C_ITS,5G-PPP}. Management of the SFN is performed by a controller, which is responsible for the optimization of transmission parameters and is the source of the safety data to be broadcast. Synchronization of transmissions is assumed to be perfect, which is enabled by a high-speed communication link with low latency. Each base station in the SFN is equipped with an antenna array with a highly directional beam. We assume that the beamwidth of the main lobe is only sufficient to cover the target cluster. In particular, we assume that the gain of the main lobe is $\mathrm{G}_{\mathrm{S,TX}}$ and the gain of the side lobes is zero. The interfering base stations are equipped with isotropic antennas with a transmission gain of $\mathrm{G}_{\mathrm{I,TX}}$.

Each vehicle in the target cluster is assumed to periodically and reliably send its location to the SFN controller via the nearest SFN base station. As such, the SFN controller can accurately estimate the geometric center of the target cluster. Each vehicle in the target cluster is equipped with an isotropic antenna with a receiving gain $\mathrm{G}_{\mathrm{RX}}$. This means that transmissions from the SFN experience both thermal noise and interference from the operator-owned network, which serves users within the city.

\begin{figure}[t]
	\centering
	\includegraphics[width=0.63\columnwidth]{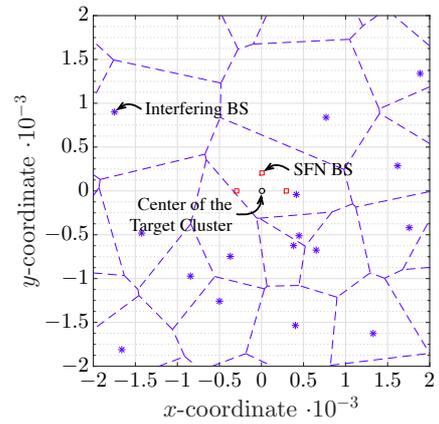}
	\caption{An instance of interfering base stations surrounding the SFN ($M = 3$, $p_\mathrm{L} = 0.2$ and $\lambda_\mathrm{I} = 10^{-4}$).}
	\label{fig.f_2}
\end{figure}

We assume that the interfering base stations are distributed throughout the city according to a two-dimensional homogeneous Poisson Point Process (PPP) $\Phi$ with intensity $\lambda_{\mathrm{I}}$, as shown in Fig.~\ref{fig.f_2}. In addition, our model distinguishes between the base stations that are in Line-of-Sight (LOS) with the center of the target cluster and those that are in Non Line-of-Sight (NLOS). As such, we assume that an interfering base station is in LOS with the center of the target cluster with probability $p_\mathrm{L}$, while the probability of an interfering base station being in NLOS is $p_\mathrm{N} = 1-p_\mathrm{L}$. Invoking the independent thinning theorem of PPPs~\cite{haenggi2013stochastic}, it follows that the PPP of the LOS interfering base stations $\Phi_{\mathrm{L}} \subseteq \Phi$ and of the NLOS base stations $\Phi_{\mathrm{N}} \subseteq \Phi$ are independent and with density $\lambda_{\mathrm{I,L}} = p_{\mathrm{L}} \lambda_{\mathrm{I}}$ and $\lambda_{\mathrm{I,N}} = p_{\mathrm{N}} \lambda_{\mathrm{I}}$, respectively. In addition, the relation $\Phi_{\mathrm{L}} \cap \Phi_{\mathrm{N}} = \emptyset$ holds. As the SFN deployment is planned, we assume that the SFN base stations are always in LOS with the center of the target cluster.

We denote the distance between the $i$-th SFN base station and the center of the target cluster as $d_{\mathrm{S},i}$. Similarly, the distance between the $j$-th interfering base station and the center of the target cluster is $d_{\mathrm{I},j}$, which is known to the SFN controller. Consider the interfering base stations; the indicator function $\mathbf{1}_{j,\mathrm{L}}$ is equal to one if base station $j$ is in LOS with respect to the center of the target cluster, and zero otherwise. The path loss component $\ell^{\mathrm{I}}(d_{\mathrm{I},j})$ associated to the $j$-th interfering base station is defined as follows
\begin{equation}
\ell^{\mathrm{(I)}}(d_{\mathrm{I},j}) = \mathbf{1}_{j,\mathrm{L}} d_{\mathrm{I},j}^{-\alpha_\mathrm{L}} + (1-\mathbf{1}_{j,\mathrm{L}}) d_{\mathrm{I},j}^{-\alpha_\mathrm{N}}\label{eq.PL}
\end{equation}
where $\alpha_\mathrm{L}$ and $\alpha_\mathrm{N}$ are the path loss exponents in the LOS and NLOS cases, respectively. The path loss component impairing the signal transmitted by the $i$-th SFN base station and received by a vehicle at the center of the target cluster simply is $\ell^{\mathrm{(S)}}(d_{\mathrm{S},i}) = d_{\mathrm{S},i}^{-\alpha_\mathrm{L}}$.

Since the interference is assumed to governed by a PPP, we exploit Slivnyak's theorem~\cite[Theorem~8.10]{haenggi2013stochastic} and assume without loss of generality that the target cluster lies at the origin $O=(0,0)$ of the axis. As such, the SINR is given by\footnote{With a slight abuse of notation, notation $\sum_{j \in \Phi}$ represents a sum across all the points in the PPP $\Phi$.}
\begin{align}
\mathrm{SINR}_O = \displaystyle\frac{\mathrm{G}_{\mathrm{S,TX}}\,\mathrm{G}_{\mathrm{RX}}\displaystyle\sum_{i=1}^M \mathrm{P}_i \, h_i \, \ell^{\mathrm{(S)}}(d_{\mathrm{S},i})}{\mathrm{W} + \mathrm{G}_{\mathrm{I,TX}}\,\mathrm{G}_{\mathrm{RX}} \, \mathrm{P}_\mathrm{I}\! \displaystyle\sum_{j \in \Phi} \, h_j \, \ell^{\mathrm{(I)}}(d_{\mathrm{I},i})},\label{eq.SINR}
\end{align}
where $\mathrm{P}_i$ is the instantaneous transmission power associated with the $i$-th SFN base station. The term $h_{i}$ is the small-scale fading coefficient due to base station $i$ (either a SFN or an interfering base station). In particular, we assume a Rayleigh channel model and, hence, $h_i$ follows an exponential distribution with mean equal to $1$. Finally, $\mbox{W}$ represents the thermal noise power.

A key challenge is to ensure that the SINR outages occur with a low probability. In our model, the main parameters affecting the outage probability are the instantaneous transmission powers $\{\mathrm{P}_i\}_{i = 1}^M$. As such, we now turn to deriving a closed-form expression for the SINR outage probability, which forms a basis to optimize the choice of each $\mathrm{P}_i$.

\section{SINR Outage and Rate Coverage Characterization}\label{sec:Res}
In order to provide an analytical model for characterizing the SINR outage and rate coverage probability, we prove the following proposition.
\begin{proposition}\label{prop.1}
Let $\mathrm{I} = \mathrm{G}_{\mathrm{I,TX}}\,\mathrm{G}_{\mathrm{RX}} \, \mathrm{P}_\mathrm{I}\! \sum_{j \in \Phi} \, h_j \, \ell^{\mathrm{(I)}}(d_{\mathrm{I},i})$ be the interference power at the center of the target cluster. The Laplace transform of $\mathrm{I}$ is
\begin{equation}
\mathcal{L}_\mathrm{I}(s) = \prod_{\mathrm{X} \in \{\mathrm{L,N}\}} \mathcal{L}_\mathrm{I,X}(s), \label{eq.c1.LI}
\end{equation}
where $\mathcal{L}_\mathrm{I,X}(s)$ is the Laplace transform of the fraction of the interference power caused by LOS base stations ($\mathrm{X} = \mathrm{L}$) or NLOS base stations ($\mathrm{X} = \mathrm{N}$), defined as follows for $\alpha_{\mathrm{X}} \geq 2$:
\begin{equation}
\mathcal{L}_\mathrm{I,X}(s) = \exp\left(-\frac{2 \lambda_{\mathrm{I,X}} \, \pi^2 \, (\mathrm{G}_{\mathrm{I,TX}}\,\mathrm{G}_{\mathrm{RX}}\mathrm{P}_{\mathrm{I}})^{\frac{2}{\alpha_{\mathrm{X}}}} \, s^{\frac{2}{\alpha_{\mathrm{X}}}}}{\alpha_{\mathrm{X}} \sin\left(\frac{2 \pi} {\alpha_{\mathrm{X}}}\right)}\right).\label{eq.c1.LX}
\end{equation}
\end{proposition}
\begin{IEEEproof}
We observe that~\eqref{eq.c1.LX} directly follows from~\cite[pp.~103-104]{haenggi2013stochastic}. In addition, we remark that $\Phi_{\mathrm{L}}$ and $\Phi_{\mathrm{N}}$ are independent PPPs. Thus, the interference determined by LOS base stations is statistically independent from the interference determined by NLOS base stations. Hence,~\eqref{eq.c1.LI} holds.
\end{IEEEproof}

The general framework for evaluating the SINR outage probability is given in the following result.
\begin{theorem}\label{th.1}
Let $\mu_{i}$ be equal to $\mathrm{P}_i \, d_{\mathrm{S},i}^{-\alpha_{\mathrm{S}}}$, for $i = 1, \ldots, M$. Consider the set $\mathcal{R} = \left\{\mu_{1}, \ldots, \mu_{M}\right\}$ and define \mbox{$\mathcal{A} = \{\mu_1, \ldots, \mu_a\}$} as the set composed by the same elements as in $\mathcal{R}$ but with no repetitions. Furthermore, we define $o_k$ as the number of repetitions of $\mu_k$ in $\mathcal{R}$, for $k = 1, \ldots, a$.
Let $\mathrm{P}_\mathrm{T}(\theta)$ to be the SINR outage probability with respect to a threshold $\theta$, i.e., the probability that $\mathrm{SINR}_O$ is smaller than a threshold $\theta$. Then, $\mathrm{P}_\mathrm{T}(\theta)$ is given by
\begin{equation}
\hspace{-0.3mm}\mathrm{P}_\mathrm{T}(\theta) =\! \prod_{j = 1}^a \mu_j^{-o_k} \! \sum_{k = 1}^a \sum_{\ell = 1}^{o_k} \mu_k^{o_k - \ell}\frac{\Psi_{k,\ell}\left(-\mu_k^{-1}\right) \Omega_{k,\ell}\left(\mu_{k}^{-1}\right)}{(o_k - \ell)! (\ell - 1)!},\hspace{-2.5mm}\label{th1.eq1}
\end{equation}
where
\begin{equation}
\Psi_{k,\ell}\left(t\right) = - \frac{\partial^{\ell - 1}}{\partial t^{\ell - 1}} \left\{\frac{1}{t}\prod_{j = 1, j \neq k}^a \left(\frac{1}{\mu_j} + t\right)^{-o_j} \right\}
\end{equation}
and
\setlength{\arraycolsep}{0.0em}\begin{eqnarray}
\Omega_{k,\ell}\left(t\right) &{}={}& (-1)^{o_k - \ell} \frac{\partial^{o_k-\ell}}{\partial x^{o_k-\ell}} \Bigg\{ e^{-\frac{\mu_k^{-1} \theta W}{\mathrm{G}_{\mathrm{S,TX}}\,\mathrm{G}_{\mathrm{RX}}} x}\notag\\
&& \cdot \, \mathcal{L}_{\mathrm{I}}\left(\frac{\mu_k^{-1} \theta \mathrm{I}}{\mathrm{G}_{\mathrm{S,TX}}\,\mathrm{G}_{\mathrm{RX}}} x\right)\Bigg\}\Bigg|_{x = 1}.
\end{eqnarray}
\end{theorem}
\begin{IEEEproof}
Consider~\eqref{eq.SINR}, $\mathrm{P}_\mathrm{T}(\theta)$ can be expressed as follows:
\begin{equation}
\mathrm{P}_\mathrm{T}(\theta) = \mathbb{P}\left[\displaystyle\sum_{i=1}^M \mathrm{P}_i \, h_i \, d_{\mathrm{S},i}^{-\alpha_\mathrm{L}} > \theta \frac{\mathrm{W} + \mathrm{I}}{\mathrm{G}_{\mathrm{S,TX}}\,\mathrm{G}_{\mathrm{RX}}} \right].\label{eq.th1.SINR}
\end{equation}
Since $\{h_i\}_{i = 1}^M$ are independent random variables following an exponential distribution with mean equal to $1$, it follows that $\{\mathrm{P}_i \, h_{i} \, d_{\mathrm{S},i}^{-\alpha_{\mathrm{S}}}\}_{i = 1}^M$ are independent exponentially distributed random variables with mean $\mu_i = \mathrm{P}_i \, d_{\mathrm{S},i}^{-\alpha_{\mathrm{S}}}$. From~\cite{693785}, it follows that the cumulative distribution function of a sum of exponentially distributed random variables, evaluated in a point $z \geq 0$, which can be expressed as follows:
\begin{equation}
\hspace{-0.1mm}\mathrm{F}(z) =\! \prod_{j = 1}^a \mu_j^{-o_k} \! \sum_{k = 1}^a \sum_{\ell = 1}^{o_k} \mu_k^{o_k - \ell}\frac{\Psi_{k,\ell}\left(-\mu_k^{-1}\right) z^{o_k - \ell} e^{-z/\mu_k}}{(o_k - \ell)! (\ell - 1)!}.\hspace{-3mm}\label{th1.proof.2}
\end{equation}
From~\eqref{th1.proof.2}, we rewrite~\eqref{eq.th1.SINR} as follows\footnote{With $\mathbb{E}_{\mathrm{X}}[\cdot]$ we refer to the expectation evaluated with respect to the random variable $\mathrm{X}$.}:
\setlength{\arraycolsep}{0.0em}
\begin{eqnarray}
\mathrm{P}_\mathrm{T}(\theta)&{}={}&\mathbb{E}_{\mathrm{I}} \left[ \mathrm{F}\left(\theta \frac{\mathrm{W} + \mathrm{I}}{\mathrm{G}_{\mathrm{S,TX}}\,\mathrm{G}_{\mathrm{RX}}}\right) \right]\notag\\
&\stackrel{(a)}{=}&\prod_{j = 1}^a \mu_j^{-o_k} \! \sum_{k = 1}^a \sum_{\ell = 1}^{o_k} \mu_k^{o_k - \ell}\frac{\Psi_{k,\ell}\left(-\mu_k^{-1}\right)}{(o_k - \ell)! (\ell - 1)!} \cdot\notag\\
&&\mu_k^{o_k - \ell} \mathbb{E}_{\mathrm{I}_0}\left[\mathrm{U}^{o_k - \ell} e^{-\mathrm{U}}\right]\label{th1.proof.3}
\end{eqnarray}
where in equality $(a)$ is the result of the variable substitution $\mathrm{U} \leftarrow \mu_k^{-1} \theta \frac{\mathrm{W} + \mathrm{I}}{\mathrm{G}_{\mathrm{S,TX}}\,\mathrm{G}_{\mathrm{RX}}}$. We observe that the following equality chain holds:
\begin{eqnarray}
\mathbb{E}_{\mathrm{I}}\left[\mathrm{U}^{o_k - \ell} e^{-\mathrm{U}}\right]&{}={}&\mathbb{E}_{\mathrm{U}}\left[\mathrm{U}^{o_k - \ell} e^{-\mathrm{U}}\right]\notag\\
&=& (-1)^{o_k - \ell} \frac{\partial^{o_k-\ell}}{\partial x^{o_k-\ell}} \left\{ \mathcal{L}_{\mathrm{U}}\left(x\right) \right\}\Bigg|_{x = 1}.\label{th1.proof.4}
\end{eqnarray}
We observe that $\mathcal{L}_{\mathrm{U}}(x)$ can be expressed in terms of the Laplace transform of $\mathrm{I}_0$ as follows:
\begin{eqnarray}
\mathcal{L}_{\mathrm{U}}(x)&{}={}&\mathbb{E}_{\mathrm{U}}\left[e^{-\mathrm{U}x}\right]\notag\\
&=& e^{-\frac{\mu_k^{-1} \theta \mathrm{W}}{\mathrm{G}_{\mathrm{S,TX}}\,\mathrm{G}_{\mathrm{RX}}} x} \mathcal{L}_{\mathrm{I}}\left(\frac{\mu_k^{-1} \theta \mathrm{I}}{\mathrm{G}_{\mathrm{S,TX}}\,\mathrm{G}_{\mathrm{RX}}} x\right).\label{th1.proof.5}
\end{eqnarray}
For these reasons, by substituting~\eqref{th1.proof.5} in~\eqref{th1.proof.4} and~\eqref{th1.proof.4} in~\eqref{th1.proof.3}, and from Proposition~\ref{prop.1} we obtain~\eqref{th1.eq1}.
\end{IEEEproof}

From Theorem~\ref{th.1}, we also have the following corollary.
\begin{corollary}\label{cl.1}
If the carnality of set $\mathcal{A}$ is equal to $M$, i.e., if $\mathrm{P}_i \, d_{\mathrm{S},i}^{-\alpha_{\mathrm{S}}} \neq \mathrm{P}_j \, d_{\mathrm{S},j}^{-\alpha_{\mathrm{S}}}$, for any $i \neq j$ and $i,j = 1, \ldots, M$, the outage probability $\mathrm{P_T}(\theta)$ can be expressed as follows:
\begin{equation}
\mathrm{P_T}(\theta) = -\prod_{j = 1}^M \mu^{-1}_j \sum_{k = 1}^M \Hat{\mathrm{D}}_k \, \mathcal{L}_{\mathrm{I}}\left(\frac{\mu_k^{-1} \theta \mathrm{I}}{\mathrm{G}_{\mathrm{S,TX}}\,\mathrm{G}_{\mathrm{RX}}}\right),\label{eq.cl.eq1}
\end{equation}
where
\begin{equation}
\Hat{\mathrm{D}}_k = \mu_k \, e^{-\frac{\mu_k^{-1} \theta \mathrm{W}}{\mathrm{G}_{\mathrm{S,TX}}\,\mathrm{G}_{\mathrm{RX}}}} \prod_{j = 1, j \neq k}^a \left(\frac{1}{\mu_j} - \frac{1}{\mu_k}\right)^{-1}.
\end{equation}
\end{corollary}
\begin{IEEEproof}
The proof arises by setting $a$ equal to $M$ and $o_k$ equal to $1$ in~\eqref{th1.eq1}, for $k = 1, \ldots, M$.
\end{IEEEproof}
\begin{remark}\label{rem1}
We observe that Theorem~\ref{th.1} and Corollary~\ref{cl.1} hold when the network of the interfering base stations has a radius that tends to infinity~\cite{haenggi2013stochastic}. When the radius is finite, the value of $\mathrm{P}_\mathrm{T}(\theta)$ is upper-bounded by the right hand side term of~\eqref{th1.eq1} or~\eqref{eq.cl.eq1}.
\end{remark}

Theorem~\ref{th.1} provides a basis to also characterize the rate outage probability; i.e., the probability that an emergency user at the center of the target cluster experiences a rate that is greater than a given threshold.
\begin{theorem}\label{th2}
The rate coverage probability as a function of a target rate value $\kappa$ is given by
\begin{equation}
\mathrm{R_C}(\kappa) = 1 - \mathrm{P_T}\left( \frac{2^{\frac{\kappa}{\mathrm{H}\sigma}} - 1}{\mathrm{J}} \right),\label{eq.th2}
\end{equation}
where $\sigma$ is the system bandwidth, while parameters $\mathrm{H}$ and $\mathrm{J}$ are rate correcting factors determined by the underlying modulation and coding scheme, as defined in~\cite{7248321}.
\end{theorem}
\begin{IEEEproof}
We refer to Shannon's rate expression \mbox{$\mathrm{H} \cdot \sigma \log_2(1 + \mathrm{J} \cdot \mathrm{SINR})$}. From~\cite[Theorem 1]{6497002} and by using Theorem~\ref{th.1}, we can express $\mathrm{R_C}(\kappa)$ as follows:
\setlength{\arraycolsep}{0.0em}
\begin{eqnarray}
\mathrm{R_C}(\kappa)&{}={}&\mathbb{P}[ \mathrm{H} \sigma \log_2(1 + \mathrm{J} \cdot \mathrm{SINR}_O) > \kappa]\notag\\
&=& \mathbb{P}\left[ \mathrm{SINR}_O > \frac{2^{\frac{\kappa}{H\sigma}} - 1}{J} \right],
\end{eqnarray}
thus,~\eqref{eq.th2} holds.
\end{IEEEproof}

By following the same reasoning as in Remark~\ref{rem1}, we observe that the right hand side of~\eqref{eq.th2} represents a lower bound for $\mathrm{R_C}(\kappa)$, in the case the radius of the interfering network is not infinite.

\subsection{Proposed Power Allocation Model for SFN}\label{subsec.PA}
By exploiting Theorem~\ref{th.1}, we now optimize the transmission power of each SFN base station. In particular, we consider the following Power Allocation (PA) problem
\vspace{-2mm}\begin{align}
	\text{(PA)} &  \quad  \min_{\mathrm{P}_1, \ldots, \mathrm{P}_M} \,\,  \sum_{i=1}^{M} \mathrm{P}_i \label{PAM.of}\\
    \text{subject to} &   \quad \mathrm{P_T}(\Hat{\theta}) \leq \Hat{\mathrm{T}}\label{PAM.c1}\\
                      &   \quad 0 \leq \mathrm{P}_i \leq \Hat{\mathrm{P}} \quad\quad\quad\quad\, \text{$i = 1, \ldots, M$}.\label{PAM.c2}
\end{align}
The objective function of the model is given by the sum of the transmission powers of each SFN base station, which we aim to minimize in~\eqref{PAM.of}. Constraint~\eqref{PAM.c1} ensures that the SINR outage probability is smaller than or equal to a target value \mbox{$\Hat{\mathrm{T}} \in [0, 1]$}, for a target SINR threshold $\Hat{\theta}$. In addition, constraint~\eqref{PAM.c2} ensures that the transmission power of each SFN base station never exceeds the maximum power level $\Hat{\mathrm{P}}$. We observe that constraint~\eqref{PAM.c2} allows $\mathrm{P}_i$ to be equal to \SI{0}{\watt} - thus allowing to switch one or more SFN base stations off if their contribution is not needed to meet the constraint~\eqref{PAM.c1}.

Due to space limitations, we are unable to provide a detailed procedure to solve the PA model. We note however that problem~\eqref{PAM.of}-\eqref{PAM.c2} is related to problem~\cite[Eq. (21)-(24)]{jsacTassi}. Thus, an heuristic solution of the PA model can be obtained by resorting to the same water-filling strategy as \mbox{in~\cite[Procedure 2]{jsacTassi}}.

If we consider typical urban vehicle speeds, the center of the target cluster is likely to move at a relatively low speed. We also observe that~\eqref{PAM.c1} is the only non-linear constraint of the PA model, whereas both~\eqref{PAM.of} and~\eqref{PAM.c2} are obliviously linear. Moreover, we note that $\mathrm{P_T}$ is a non-increasing function with respect to $\{\mathrm{P}_i\}_{i = 1}^M$. For these reasons, it is also possible to efficiently derive an heuristic solution to the PA model by means of a genetic approach~\cite{jsacTassi}.

\section{Numerical Results}\label{sec:NR}
In this section, we numerically investigate our emergency SFN setup. In particular, we detail our simulation framework and validate our theoretical model against Monte Carlo simulation. Finally, we investigate how the proposed PA model performs in different simulation scenarios.

As noted in Remark~\ref{rem1}, relations~\eqref{th1.eq1} and~\eqref{eq.th2} holds. for networks of interfering base stations with a radius $R$ that tends to infinity, which clearly is not possible to simulate. In particular, as the radius of the network reduces, the divergence of the equations in~\eqref{th1.eq1} and~\eqref{eq.th2} increases from the actual SINR outage and rate coverage probability values. From~\cite[Eq.~(3.5)]{Haenggi2008}, we observe that the simulation accuracy error $\epsilon$ is $R \geq \epsilon^{-1/(\alpha_\mathrm{L} - 1)}$. As such, we considered $\alpha_\mathrm{L} = 2.5$ and simulated scenarios with radius $R = \SI{1000}{\meter}$ - thus ensuring a value of $\epsilon$ not smaller than $3.16 \cdot 10^{-5}$. We account for the possibility of an interfering base station to be in NLOS with respect to the center of the target cluster by setting $p_\mathrm{N} = 0.8$. Furthermore, we set $\alpha_\mathrm{N} = 3.5$~\cite{TR_36_814}.

\begin{figure}[tb]
\centering
\subfloat[$\mathrm{P_T}(\theta)$ as a function of $\theta$, for different values of $\lambda_{\mathrm{BS}}$.]{\label{fig.f_v1.1}
	\includegraphics[width=1\columnwidth]{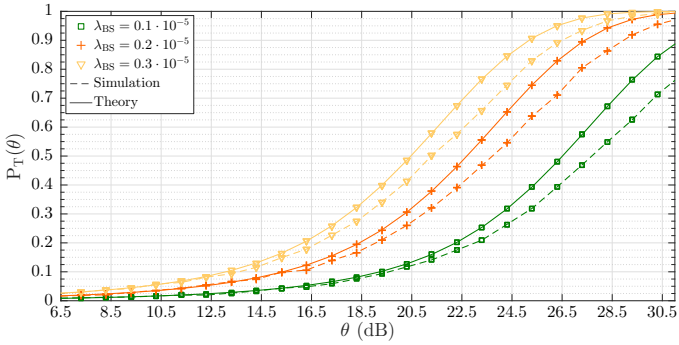}
}\\
\subfloat[$\mathrm{P_T}$ as a function of $\lambda_{\mathrm{BS}}$, for different values of $\theta$.]{\label{fig.f_v1.2}
	\includegraphics[width=1\columnwidth]{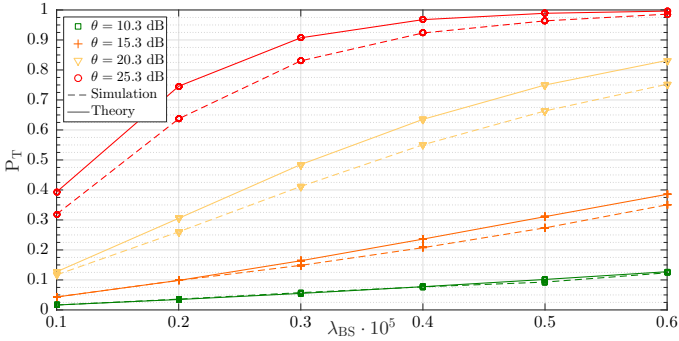}
}
\caption{Comparison of simulated and theoretical SINR outage probability $\mathrm{P_T}$.}\vspace{-4mm}
\label{fig.f_v1}
\end{figure}

We considered a system bandwidth $\sigma = \SI{50}{\mega\hertz}$, which is the equivalent bandwidth of an LTE-A system employing $50$ Resource Block Pairs for downlink communications on each eMBMS subframe~\cite{jsacTassi}. We set the thermal noise power $\mathrm{W}$ equal to \mbox{$10 \log_{10}(k \cdot T \cdot \sigma \cdot 10^3)$ \SI{}{\dBm}}, where $k$ is the Boltzmann constant and the temperature $T$ is set equal to \SI{290}{\kelvin}~\cite{rappaport2014millimeter}. Furthermore, we set the vehicle reception antenna gain $\mathrm{G_{RX}} = \SI{10}{\dB}$ and the transmission antenna gain associated with the main lobe of each SFN base station $\mathrm{G_{S,TX}} = \SI{20}{\dB}$, whereas the transmission antenna gain of each interfering base station $\mathrm{G_{I,TX}}$ has been set equal to $\SI{7}{\dB}$. Finally, we remark that the goal of our theoretical framework (see Section~\ref{sec:Res}) is that of characterizing the SINR outage and a rate coverage of an ideal emergency user positioned at the center of the target cluster, the antenna beamwidth of each SFN base station has no impact on both our theoretical and simulation results.

In the reminder of the section, we consider a scenario with $M = 3$ SFN base stations located at the coordinates $\{(-300, 0), (300, 0), (0, 200)\}$\SI{}{\meter} and characterized by a maximum instantaneous transmission power $\Hat{\mathrm{P}}$ equal to $\SI{30}{\watt}$. The instantaneous transmission power of each interfering base station $\mathrm{P_I}$ is set equal to $\SI{10}{\watt}$.

\begin{figure}[t]
	\centering
	\includegraphics[width=1\columnwidth]{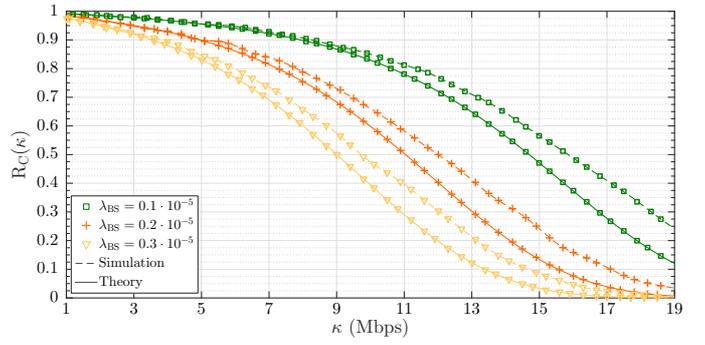}
	\caption{Comparison of simulated and theoretical rate coverage probability $\mathrm{R_C}(\kappa)$ as a function of $\kappa$, for different values of $\lambda_{\mathrm{BS}}$.}\vspace{-5mm}
	\label{fig.f_v2}
\end{figure}

Fig.~\ref{fig.f_v1} compares the simulated and theoretical SINR outage probability given in \eqref{th1.eq1}. In particular, Fig.~\ref{fig.f_v1.1} shows the SINR outage probability as a function of $\theta$, for different densities of interfering base stations \mbox{$\lambda_\mathrm{BS} = \{0.1 \cdot 10^{-5}, 0.2 \cdot 10^{-5}, 0.3 \cdot 10^{-5}\}$}. As expected (see Remark~\ref{rem1}), \eqref{th1.eq1} forms an upper-bound for the corresponding simulation results. In particular, the proposed theoretical model, as in~\eqref{th1.eq1}, appears to be quite tight for values of $\theta$ that are smaller than or equal to $\SI{16.5}{\dB}$. For larger values of $\theta$, the absolute error between simulation and theory increases but never exceeds $0.13$. Fig.~\ref{fig.f_v1.2} provides the same comparison as in Fig.~\ref{fig.f_v1.1} but in this case $\mathrm{P_T}$ is expressed as a function of $\lambda_\mathrm{BS}$, for different values of $\theta$. Although the absolute error between simulation and theory increases as $\lambda_\mathrm{BS}$ and $\theta$ increase, we observe that the proposed SINR outage probability model remains proves to be a tight upper-bound.

By considering the same simulation parameters as in Fig.~\ref{fig.f_v1}, Fig.~\ref{fig.f_v2} compares simulated rate coverage probability values against those obtained from~\eqref{eq.th2} as a function of $\kappa$. In this performance investigation, we considered an LTE-A communication system. As such, we set parameters $\mbox{H}$ and $\mbox{J}$ of~\eqref{eq.th2} equal to $0.17$ and $0.06$, respectively~\cite{7248321}. Given that the expression of $\mathrm{R_C}(\kappa)$ follows from $\mathrm{P_T}(\theta)$, as expected we observe that the reported theoretical results lower-bound the simulated ones. Also in this case, the proposed lower-bound appears to be tight as the maximum gap between simulations and theory is smaller than $0.13$.

\begin{figure}[tb]
\centering
\subfloat[PA model performance as a function of $\Hat{\theta}$, for different values of $\lambda_{\mathrm{BS}}$.]{\label{fig.f_v3.1}
	\includegraphics[width=1\columnwidth]{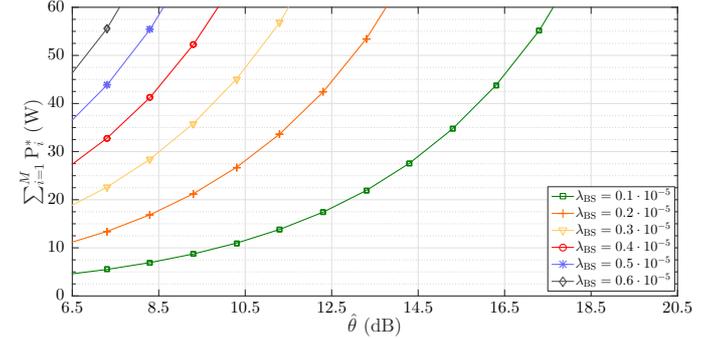}
}\\
\subfloat[PA model performance as a function of $\lambda_{\mathrm{BS}}$, for different values of $\Hat{\theta}$.]{\label{fig.f_v3.2}
	\includegraphics[width=1\columnwidth]{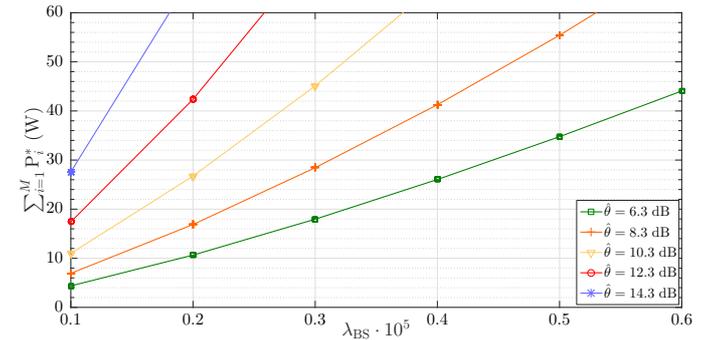}
}
\caption{Optimum value of $\sum_{i=1}^M \mathrm{P}^*_i$.}\vspace{-3mm}
\label{fig.f_v3}
\end{figure}

We regard $\{\mathrm{P}^*_i\}_{i=1}^M$ as the optimized values of $\{\mathrm{P}_i\}_{i=1}^M$ obtained by solving the proposed PA model by resorting to a genetic strategy (see Section~\ref{subsec.PA}). In particular, Fig.~\ref{fig.f_v3} shows the optimized values of the objective function~\eqref{PAM.of}, i.e., $\sum_{i=1}^M \mathrm{P}^*_i$. In particular, we refer to a scenario where the SINR outage probability threshold $\mathrm{\Hat{T}}$ is set equal to $0.1$.

Consider Fig.~\ref{fig.f_v3.1}, it shows optimized values of the objective function~\eqref{PAM.of} as a function of the SINR threshold $\Hat{\theta}$, for different values of $\lambda_\mathrm{BS}$. As expected, we observe that $\sum_{i=1}^M \mathrm{P}^*_i$ increases as the density of the  interfering base stations, or as the value of the target SINR threshold, increase. However, the considered optimized PA solution allows us to significantly reduce the amount of total instantaneous transmission power compared to a static PA solution - enforcing $\mathrm{P}_i = \mathrm{\Hat{P}}$ and, hence, leading to a total SFN transmission power equal to $M \cdot \mathrm{\Hat{P}} = \SI{90}{\watt}$. This is particularly evident for $\Hat{\theta} = \SI{6.5}{\dB}$ and $\lambda_\mathrm{BS} = 0.1 \cdot 10^{-5}$, where our PA solution achieves a total SFN transmission power that is more than $20$ times smaller the static PA solution. The same conclusions can be drawn from Fig.~\ref{fig.f_v3.2}, which slows $\sum_{i=1}^M \mathrm{P}^*_i$ as a function of different densities of interfering base stations, for different values of $\Hat{\theta}$.

\section{Conclusions}\label{sec:CL}
At present, the IEEE 802.11p/DSRC and LTE-A standards cannot entirely support the reliable transmission of information in vehicular emergencies as required by next-generation ITS services. In this paper, we have proposed that \mbox{municipality-owned} SFNs are a promising alternative means of ensuring that drivers have access to this information. To evaluate these SFNs, we have obtained performance guarantees of these networks in terms of bounds on outage probabilities using techniques from stochastic geometry. These bounds form a basis for optimizing the power allocation of each base station in the SFN, which is important when these base stations rely on off-grid power sources. In the considered scenarios, we have shown that the proposed PA model can ensure and overall transmission power footprint that: (i) can be up to $20$ times smaller than a static PA solution, and (ii) meets target SINR outage constraints.

\bibliographystyle{IEEEtran}
\vspace{-2.4mm}\bibliography{bib}
\end{document}